\documentclass{icrc29}

\usepackage{graphicx,amssymb,amsmath,times}

\begin{document}

\title[Lateral width of shower image in the Auger fluorescence
  detector.]{Lateral width of shower image in the Auger fluorescence
  detector.}

\author[H. Barbosa et al.] {H. Barbosa$^a$, V. de Souza$^b$,
  C. Dobrigkeit$^a$, R. Engel$^c$, D. G\'ora$^d$, D. Heck$^c$, 
\newauthor P. Homola$^d$, H. Klages$^c$, 
G. Medina-Tanco$^b$, J.A. Ortiz$^b$, J. P\c{e}kala$^d$,
\newauthor M. Risse$^c$,
B. Wilczy\'nska$^d$, and  H. Wilczy\'nski$^d$ for the Pierre Auger
Collaboration \\
}

\presenter{Presenter: V. de Souza (vitor@astro.iag.usp.br)}

\maketitle

\begin{abstract}
     
The impact of the lateral distribution of light in extensive air showers on the detection 
and reconstruction of shower profiles is investigated for the Auger fluorescence telescopes.
Based on three-dimensional simulations, the capability of the Auger telescopes to measure
the lateral distribution of light is evaluated. The ability to infer the actual lateral 
distribution is confirmed by the comparison of detailed simulations with real data. 
The contribution of pixels located far from the axis of the shower image
is calculated and the accepted signal is rescaled in order to reconstruct
a correct shower profile. The analysis presented here shows that:
(a) the Auger telescopes are able to observe the lateral distribution of showers
and (b) the energy corrections to account for the signal in outlying pixels
can exceed 10\%, depending on shower geometry.

\end{abstract}


\section{Introduction}
\label{introduction}

The Pierre Auger Observatory is a hybrid detector based on
fluorescence telescopes and water Cerenkov tanks \cite{1}. The basic
configuration of the Auger telescopes is a 
Schmidt camera consisting of a 1.1\,m radius aperture (including a ring
of corrector lenses), and a spherical mirror with a $30^\circ \times 30^\circ$ of field
of view. The fluorescence light is detected by an array of 440
photomultipliers, each with $1.5^\circ$ diameter field of view. The signal is
sampled in time slots of 100\,ns.  

The amount of fluorescence light produced by a shower particle is proportional
to the energy it deposits in air via ionization losses. As can be
seen in reference \cite{2}, the energy deposited by the
particles in an air shower has a wide lateral distribution. According
to that work, electrons and
positrons at distances between 100 and 1000 meters from the shower axis account for
15 to 20\% of the energy released in air by a shower.     

\begin{figure}
  \begin{center}
    \includegraphics[height=5.5cm,width=5.5cm]{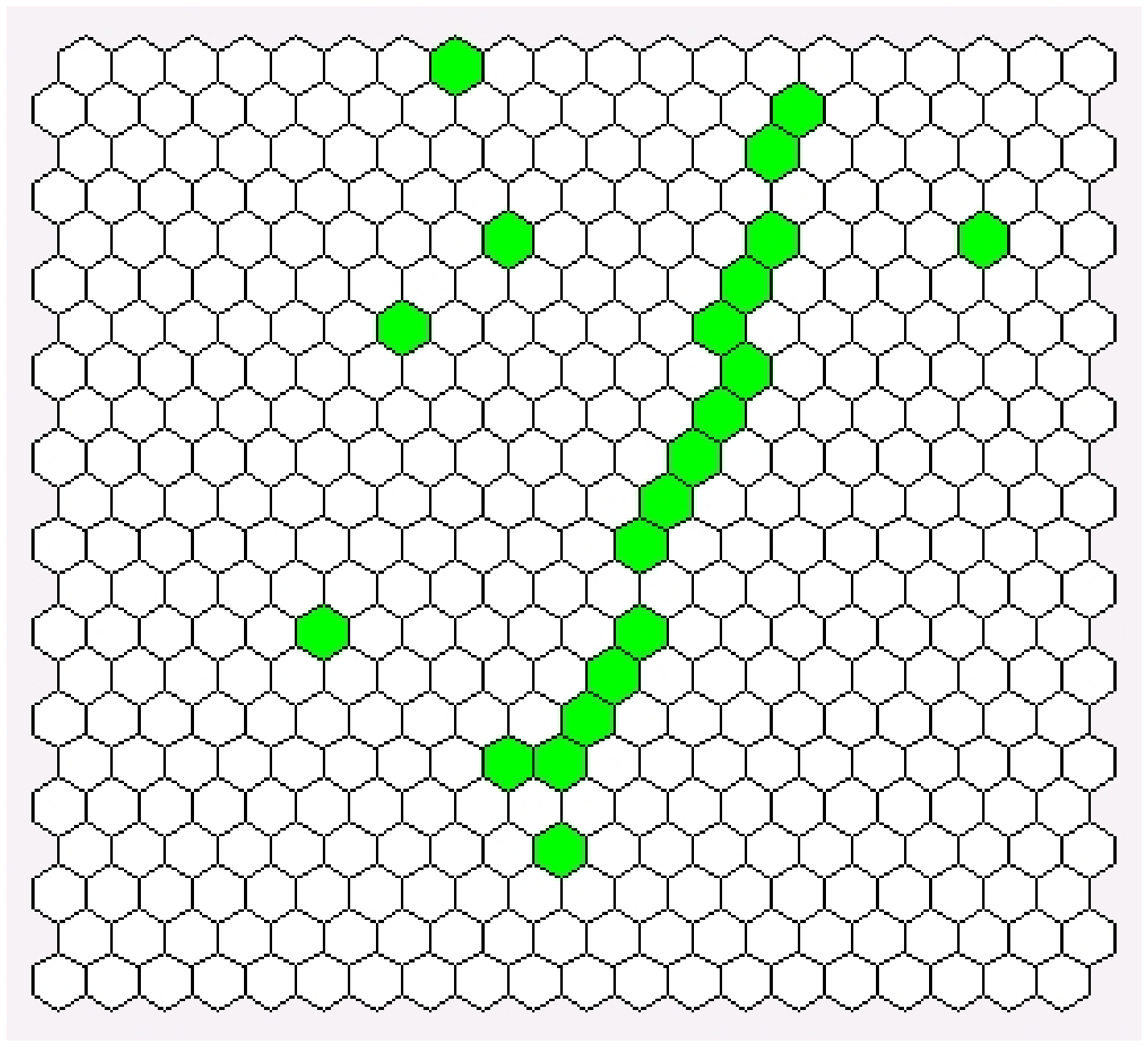} \hspace{1cm}
    \includegraphics[height=5.5cm,width=5.5cm]{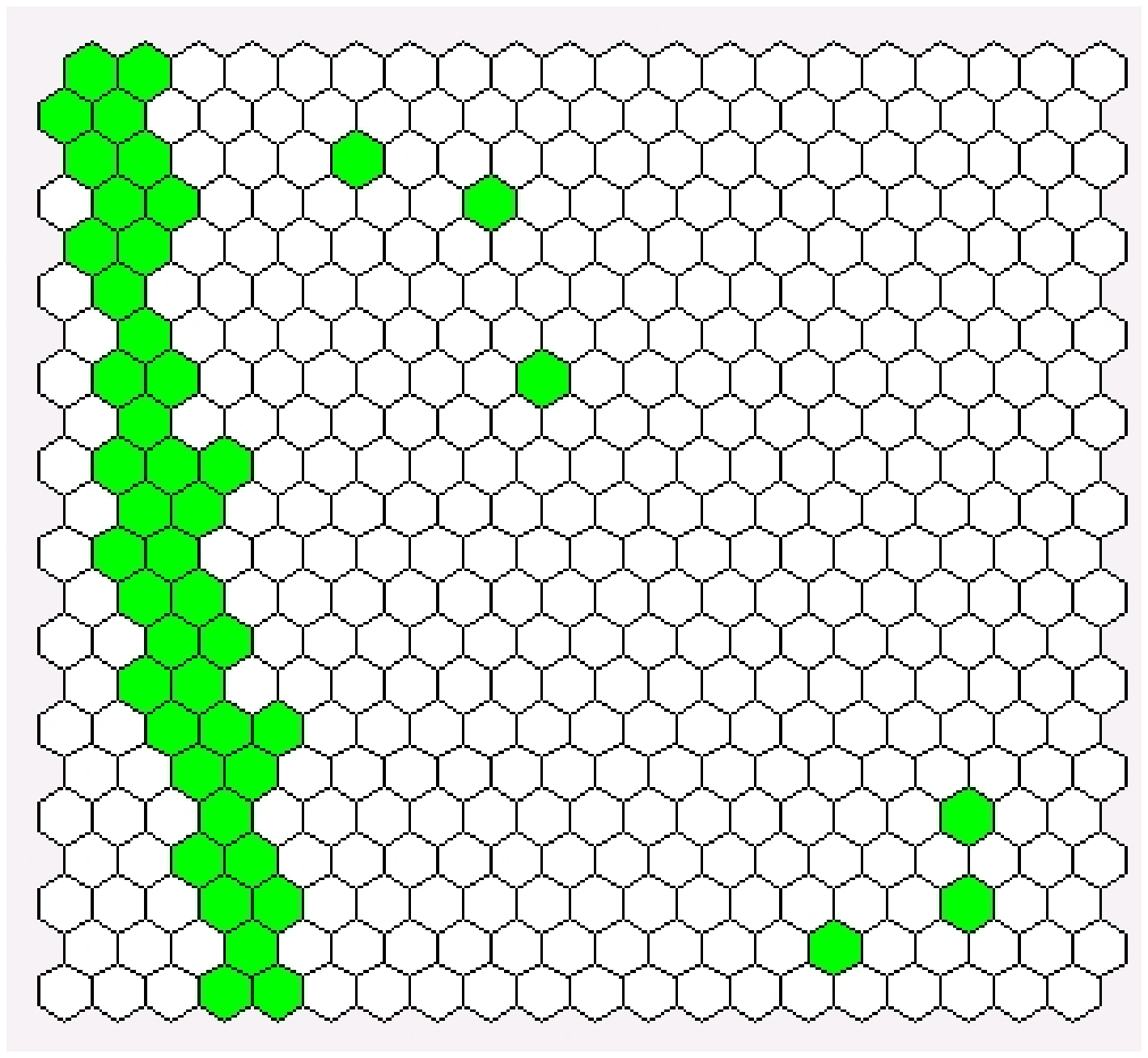}
  \end{center}
  \caption{Image of two showers in the photomultiplier camera. The
    reconstructed energy of both showers is 2.2 EeV. The shower on the left
    had a core 10.5 km from the telescope,  while that on the right landed 4.5 km away.
    Note the number of pixels and
    the lateral spread in the image in each shower.}
  \label{fig:cam}
\end{figure}

Figure \ref{fig:cam} shows examples of two events measured by the Auger
telescopes. Both events had reconstructed energies of 2.2\,EeV. 
The shower illustrated in the left panel landed with a core 10.5\,km from
the telescope while the shower illustrated in the right panel landed only
4.5\,km away. Note the difference in lateral
spread of the signal in these showers.

In this paper, we investigate how the Pierre Auger telescopes detect
the lateral distribution of particles in the shower. We compare the 
simulated lateral spread of light on the camera with that seen in real events.
Finally, a correction is proposed for the reconstructed shower energy, to take account
of the fraction of
light falling into pixels located far from the axis of the image.

\vspace{-0.2cm}

\section{Detection of the Lateral Distribution by the Auger Telescopes}

A three-dimensional shower simulation program CORSIKA \cite{3} was used 
to evaluate the 
lateral distribution of particles in the shower and consequently
the lateral spread of the signal on the camera. The energy deposited
by particles in the atmosphere as given by CORSIKA can be converted to
fluorescence photons and propagated to the telescope aperture.

The telescope simulation and the shower reconstruction  have been done
using the official Auger collaboration programs described in
references \cite{4,5}. 

In order to investigate the capability of the Auger telescopes to
measure the lateral distribution of showers, we have simulated two
sets of 100 vertical 
showers initiated by $10^{19}$ eV protons. One set was simulated with the
three-dimensional approach and the other with a one-dimensional simulation
using a Gaisser-Hillas function for the longitudinal profile.

Two parameters have been investigated in this study: the number of 
triggered pixels and the $\zeta$ angle. The $\zeta$ angle is the radius of a
circle (measured in degrees) on the photomultiplier camera
which maximizes the signal to noise ratio, $S/N$ for collected light. The distribution of
triggered pixels on the camera allows us to determine the main
track of the shower by fitting a line to the hit pixels.  Signal
from a pixel is included in the measured light flux if the pixel center
lies within an angle $\zeta$ from the track axis.  The value of  $\zeta$
is varied to search for a maximum $S/N$.  After the search, $\zeta$ is set to a fixed value for
the entire track.

\begin{figure}[t]
  \begin{center}
    \includegraphics[width=7.6cm]{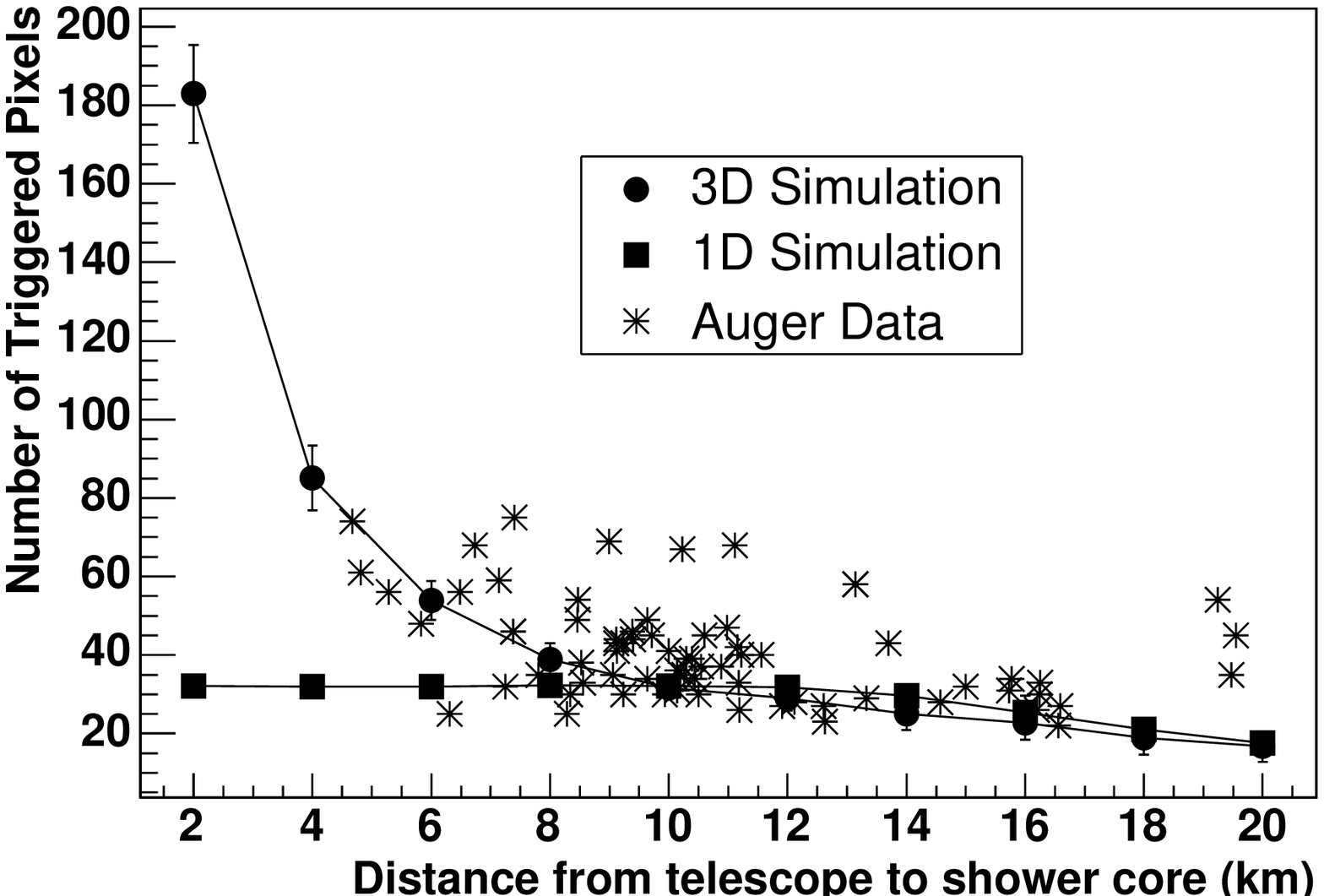}
    \includegraphics[width=7.6cm]{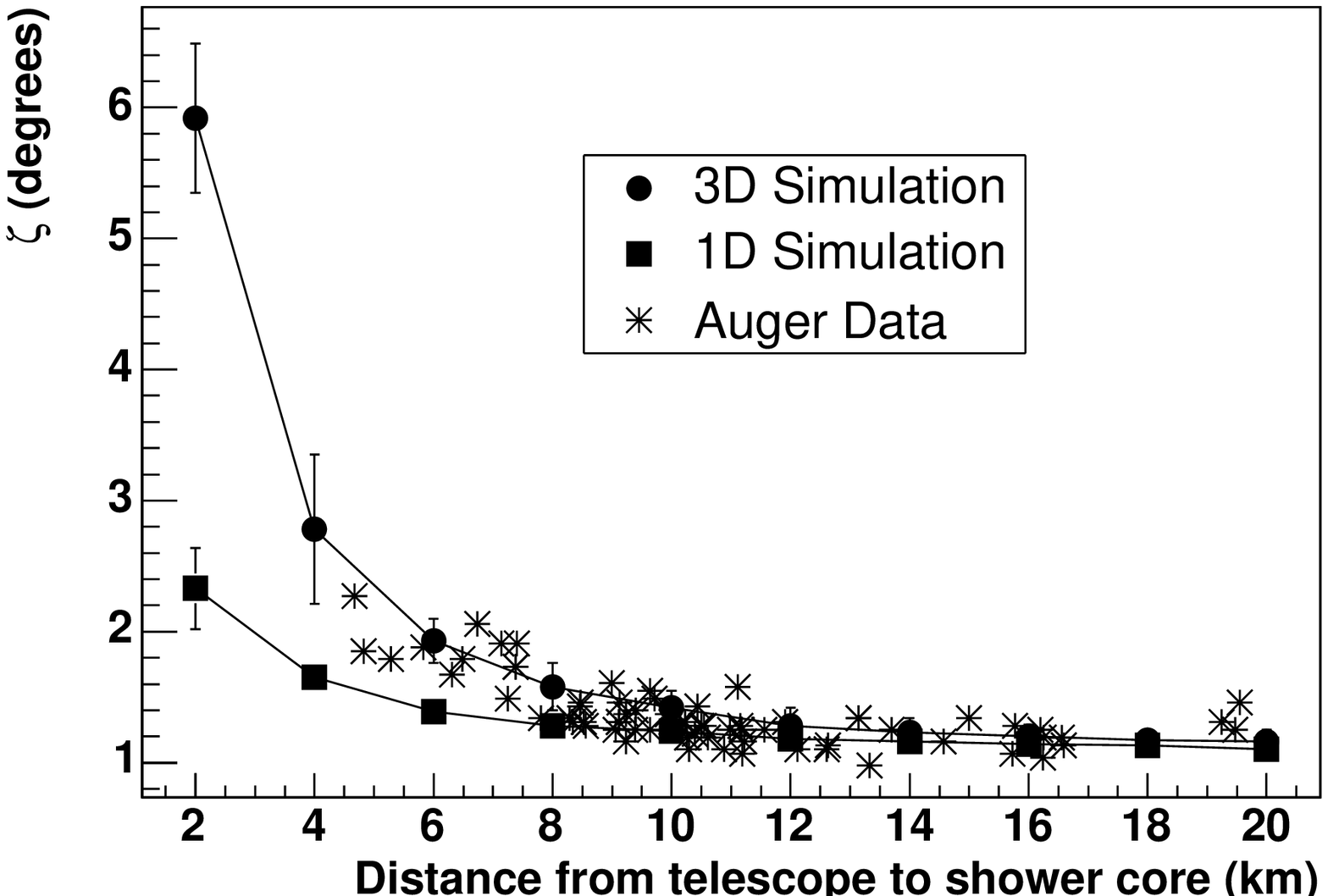}
  \end{center}
   \caption{Number of pixels and $\zeta$ as a function of the distance
  between the telescope and the shower core.}
   \label{fig:zetadist}
\end{figure}

Figure \ref{fig:zetadist} shows the expectation for the number of 
triggered pixels and $\zeta$,  obtained for one and
three-dimensional simulations as a function of the distance between the
telescope and the core position of the shower.  Figure
\ref{fig:zetadist} shows that for cores closer than 10\,km 
the lateral distribution of the particles in
the shower produces a  measurable and important spread of the signal
on the photomultiplier camera at primary energies of $10^{19}$ eV.  Real
data are also shown in the figures, from showers measured to have energy
between $10^{18.5}$ and $10^{19.5}$ eV, and these data follow the expectation
from the three-dimensional simulation.  This shows
the capability of the telescopes to measure the lateral distribution of the
signal produced by showers.

\vspace{-0.3cm}

\section{Influence of the Lateral Distribution in the Energy
  Reconstruction} 

The primary energy of the shower is calculated based on the amount of
fluorescence light recorded by the fluorescence telescope. The
standard reconstruction procedure sums the 
measured charge in each time slot (for pixels within the radius
$\zeta$) and converts it to the number of photons at the telescope
aperture using calibration constants. 

However, this method is most suitable for distant showers where the
light collected within the radius $\zeta$ corresponds to about 100\%
of the total signal. Some differences between the signal
within $\zeta$ and the total signal produced by a shower may exist for
nearby showers. 

As can be seen in reference \cite{6}, the  fraction of energy
deposited $F(r)$ within a distance $r$, measured in Moli\`ere  
units, can be well parameterized as a function of an effective shower  
age parameter only. 

We have used this parameterization to calculate the 
total signal produced by the passage of the shower, which may be
distributed among several neighboring detector pixels.
For a given geometry of the shower, we find the collected signal $L_{\zeta}(t)$
within the angular distance $\zeta$ at each time interval.  
Then, for the given $\zeta$ and the detector-to-shower distance $R_{0}$,
the effective radius around the shower axis for which the produced signal
will be accounted for in the standard energy reconstruction is given
by $r_{0} = R_{0} \tan(\zeta)$.

The fraction of energy deposited in air within $r_{0}$, and therefore the amount of
fluorescence light, can be calculated according to
reference \cite{6}.
Finally, the signal outside $r_{0}$ can be considered in the energy
reconstruction by rescaling $L_{\zeta}(t)$ according to the formula
$L_{total}(t) = L_{\zeta}(t)/F(r_0)$.

\begin{figure}[t]
  \begin{center}
    \includegraphics[width=7.5cm,angle=-90]{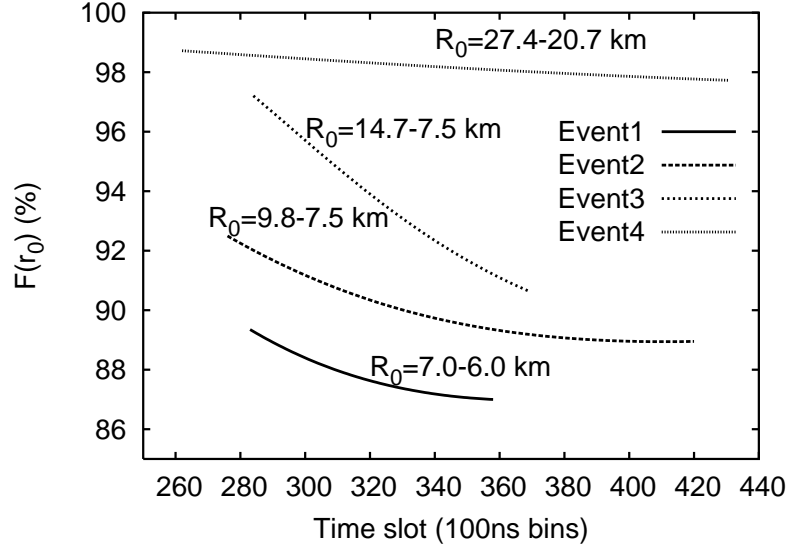}
  \end{center}
  \caption{Fraction of light collected within the angle $\zeta$ versus time for four
      events measured by the Auger telescopes.}
  \label{fig-frac}
\end{figure}

Figure \ref{fig-frac} shows four events measured by the Auger
telescopes to which this procedure has been applied.
It is seen that for Event1 $F(r_{0})$ changes from 89\%  for a
distance to the shower of $R_{0}=7.0$ km to  87\%  for $R_{0}=6.0$ km. 
Accepting only a fraction of the signal contained within
$\zeta$ directly  influences the reconstructed primary
energy of the shower. In Table \ref{tab2} we present the influence
of the correction on the Gaisser-Hillas fit to
the reconstructed number of particles in the showers. It is seen
that this correction changes both the number of particles at the
shower maximum and the position of the shower maximum. These changes 
lead to different estimates of primary energy. In the last column of
Table \ref{tab2} the relative differences 
$k_{E} = (E^{total}_{0} - E^{\zeta}_{0}) / E^{\zeta}_{0} $ are listed.
One sees that $k_{E}$ is always positive and decreases from 14\% for a  
distance to shower maximum of $R_0$=6.5 km to 2\% for $R_0$=23 km.

\begin{table}[h]                                 
\begin{center}                                                                                 
\caption{Comparison of Gaisser-Hillas function parameters based on the $L_{total}(t)$
 and $L_{\zeta}(t)$ light profiles and their influence on primary
 energy.
\label{tab2}}                                                                                                                     
\vskip 0.5cm                                                                                                                      
\begin{tabular}{ccccccccc}                                                                                                          
\hline                                                                                                                            
\hline                                                                                                                            
Event         & $R_0$    &  $N_{max}^{\zeta}$ & $N_{max}^{total}$ &  $X_{max}^{\zeta}$ &$X_{max}^{total}$&$ E_{0}^{\zeta}$& $ E_{0}^{total}$& $k_E$ \\                                                    
              & (km) &  ($10^{9})$ & (10$^{9}$) & ($g/cm^{2}$)     &($g/cm^{2}$) & (EeV)&  (EeV)& (\%) \\                                                                    
\hline                                                                                                                            
 Event1  & 6.4 & 0.93     & 1.06   & 701 & 706 & 1.370 &1.562 & 14\\
 Event2    & 8   & 6.57     & 6.88   & 759 & 767 & 9.853&10.40 & 6\\
 Event3   & 11  & 2.12     & 2.19   & 637 &642 &2.950 &3.100 &5\\
 Event4  & 23  & 12.85     & 13.10   & 752 & 753 &19.20 &19.57 & 2  \\
                                                               
\hline                                                                                                                            
\hline                                                                                                                            
\end{tabular} 
\end{center}                                                                                                                    
\end{table}                                                                                                                       

\section{Conclusion}

The Pierre Auger fluorescence telescopes are able to detect the
lateral distribution of particles in close-by showers and an energy correction
must be applied due to this effect. The correction can exceed 10\%, as
shown in Table \ref{tab2}, depending on the geometry of the shower.

\section{Acknowledgments} This work was partially supported by the Polish
Committee for Scientific Research under grants  No. PBZ KBN
054/P03/2001  and 2P03B 11024, in Germany by the DAAD under grant
No. PPP 323 and in Brazil by CNPq and FAPESP.

\end{document}